\documentclass{article}

\usepackage{arxiv}
\usepackage{subfigure}
\usepackage[utf8]{inputenc} 
\usepackage[T1]{fontenc}    
\usepackage{hyperref}       
\usepackage{url}            
\usepackage{booktabs}       
\usepackage{subcaption}
\usepackage{amsfonts}       
\usepackage{nicefrac}       
\usepackage{microtype}      
\usepackage{amsmath}
\usepackage{lipsum}         
\usepackage{graphicx}
\usepackage{natbib}
\usepackage{doi}
\usepackage{tikz} 
\usepackage{calc}

\usepackage{multirow}
\usepackage{array}

\usepackage{cleveref}       
\title{Using data assimilation tools to dissect GraphDOP}

\newbox{\orcid}\sbox{\orcid}{\includegraphics[scale=0.06]{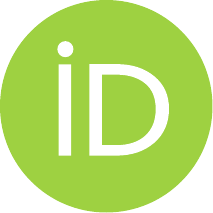}} 

\author{
    {\href{https://orcid.org/0000-0003-2808-0463}{\usebox{\orcid} Patrick Laloyaux}}
    \And {\href{https://orcid.org/0009-0007-7798-6524}{\usebox{\orcid} Mihai Alexe}}
    \And {\href{https://orcid.org/0000-0002-6070-2544}{\usebox{\orcid} Eulalie Boucher}}
    \And {\href{https://orcid.org/0000-0002-3662-5382}{\usebox{\orcid} Peter Lean}}
    \And {\href{https://orcid.org/0000-0003-1869-3426}{\usebox{\orcid} Ewan Pinnington}}
    \And {\href{https://orcid.org/0000-0003-3952-586X}{\usebox{\orcid} Simon Lang}}
    \And {\href{https://orcid.org/0000-0002-7484-3372}{\usebox{\orcid} Tobias Necker}}
    \And {\href{https://orcid.org/0000-0002-5193-3041}{\usebox{\orcid} Anthony McNally}}
    \And
    European Centre for Medium-Range Weather Forecasts (ECMWF)
}

\renewcommand{\shorttitle}{Using data assimilation tools to dissect GraphDOP}

\begin{document}

\maketitle

\begin{abstract}
The Data Assimilation (DA) community has been developing various diagnostics to understand the importance of the observing system in accurately forecasting the weather. They usually rely on the ability to compute the derivatives of the physical model output with respect to its initial condition. For example, the Forecast Sensitivity-based Observation Impact (FSOI) estimates the impact on the forecast error of each observation processed in the DA system.
This paper presents how these DA diagnostic tools are transferred to Machine Learning (ML) models, as their derivatives are readily available through automatic differentiation. We specifically explore the interpretability and explainability of the observation-driven GraphDOP model developed at the European Centre for Medium-Range Weather Forecasts (ECMWF).
The interpretability study demonstrates the effectiveness of GraphDOP's sliding attention window to learn the meteorological features present in the observation datasets and to learn the spatial relationships between different regions. Making these relationships more transparent confirms that GraphDOP captures real, physically meaningful processes, such as the movement of storm systems. 
The explainability of GraphDOP is explored by applying the FSOI tool to study the impact of the different observations on the forecast error. This inspection reveals that GraphDOP creates an internal representation of the Earth system by combining the information from conventional and satellite observations.
\end{abstract}

\section{Introduction}
Weather forecasting guides decisions that affect public safety, infrastructure, and emergency response \citep{Venuti25}. It is therefore crucial to ensure that the predictions generated by complex models can be understood, trusted, and validated by domain experts. The research community has been developing different tools over the last decades to diagnose the behaviour of weather models and the way they are initialised using Data Assimilation (DA). For example, many Numerical Weather Prediction (NWP) centres monitor the difference between observations and forecasts at different lead times to identify instrument issues or model deficiencies \citep{Dahoui24}. Observing System Experiments (OSEs) are additionally run on a regular basis, where a specific part of the observing system (e.g. microwave radiances) is withheld to quantify its impact on the DA system and on forecast skill scores \citep{Bouttier01,Bormann19}. As the range of future weather possibilities is usually captured by an ensemble of weather forecasts, monitoring the variability of the ensemble members around their mean provides complementary insights to quantify the forecast uncertainties \citep{Bonavita12,Rodwell16,Buizza19}.

The European Centre for Medium-Range Weather Forecasts (ECMWF) was a pioneer in the operational implementation of Four-Dimensional Variational (4D-Var) data assimilation; this represented a major breakthrough in numerical weather prediction \citep{Rabier00}. At its heart, 4D-Var relies on the availability of the tangent linear and adjoint versions of the Integrated Forecast System (IFS) atmospheric model, that are used to trace back forecast errors to changes in the initial conditions. The adjoint model is a key component in computing Forecast Sensitivity to Observations (FSO) diagnostics, which measure how sensitive a forecast is to individual observations \citep{Baker00}. It is also required to produce Forecast Sensitivity-based Observation Impact (FSOI) diagnostics, which quantify the actual impact of each observation on the forecast error \citep{Langland04,Cardinali09}.

Weather prediction is undergoing a transformation with data-driven models proving to be highly effective and, in many ways, surpassing the skill of physics-based counterparts (e.g. \cite{keisler2022}, \cite{lam2023}, \cite{lang2024}, \cite{Price2025GenCast}, \cite{Lang2024AIFSCRPS}). These models have primarily been trained on reanalysis datasets, notably ERA5 \citep{hersbach2020era5}. In contrast, the AI-Direct Observation Prediction (AI-DOP) approach proposed in \cite{mcnally2024} aims to learn directly from observational data without relying on prior physical knowledge or numerical modelling output. This global end-to-end weather prediction model trained exclusively from observations was implemented using Graph Neural Networks (GNNs) in GraphDOP \cite{alexe2024}. Learning directly from observational data has been also investigated to some extent by other research groups \citep{Allen2025Aardvark,vandal2024,keller2024,sun2024}.  

\citet{Lean25} and \citet{Boucher25} showed that GraphDOP creates a coherent internal representation of the Earth system and of the physical processes that the observations are sensitive to, in a similar way to standard NWP models, which are built on well-established physical principles like the laws of thermodynamics and fluid dynamics. \textcolor{black}{This paper demonstrates how sensitivity-based diagnostics that have been initially developed in DA with standard NWP models can be applied to Machine Learning (ML) models.} Such tools allow to explore the interpretability and explainability of the GraphDOP model, confirming that it captures real and physically meaningful processes. The paper aims to make GraphDOP more understandable by analysing its internal representation. This allows meteorologists to assess that the model's reasoning aligns with established atmospheric science. It not only fosters trust and transparency but also enables collaboration between ML experts and meteorologists, ensuring that complex models are used effectively in operational settings. This article is organized as follows. In section~2, a brief overview of GraphDOP summarises how the encoders, processor and decoders are implemented. The interpretability of GraphDOP is discussed in Section~3 where different mechanisms of its structure are illustrated. Section~4 covers the explainability of GraphDOP where the FSOI tool is implemented to study the impact of different observations on the forecast error. Finally, conclusions and future directions of research are discussed in Section~5.

\section{The GraphDOP model}
GraphDOP is a data-driven weather forecasting system that learns directly from Earth system observations, including satellite radiances and in-situ measurements, to produce weather forecasts \citep{alexe2024}. This represents a significant departure from traditional NWP systems or other ML models, as GraphDOP does not require input from conventional reanalyses or model state estimates.
 
GraphDOP is built around an encoder-processor-decoder architecture, as shown, e.g., in Figure 1 of \cite{Lean25}. The encoders ingest the observational input data and create a latent representation of the atmospheric state by aggregating information from the observations across space and time. The model used in this study has 40320 latent space nodes, each with 1024 features distributed on an o96 octahedral reduced Gaussian grid \citep{malardel2016} with a spatial resolution of approximately 1 degree (110 km). The latent representation is then passed through the processor, which is the module responsible for evolving the latent state forward in time. Its role is similar in concept to the physical dynamical core of a traditional numerical weather model (like ECMWF's IFS). The processor is implemented as a sliding window attention transformer across latitude bands (cf. Figure 2 in \cite{lang2024}), while the encoders and decoders use graph attention \citep{alexe2024,Lean25}. Finally, the decoders project the latent Earth system state representation back to physical quantities (e.g., radiances and conventional observations); it plays a role similar to that of an observation operator in a traditional (physics-based) DA system. The use of GNN-based encoders and decoders means that input observations need not be gridded before being passed to the system. During inference (forecasting), GraphDOP can produce predictions at arbitrary times and locations inside the target window, including at locations/times where "real" observation data are not available. This allows GraphDOP to produce gridded forecasts when presented with just the metadata associated with the target observation locations, e.g., time, spatial coordinates, station or pressure altitude, solar zenith angle, and instrument viewing angle, where applicable. The training objective used here is a weighted mean square error (WMSE) loss similar to that employed by \citet{alexe2024}.

To provide the best possible forecast skill scores, GraphDOP was originally trained on a wide variety of observing networks, including radiances and conventional observations (see Table~1 in \citet{alexe2024}). As the focus of this paper is to diagnose the GraphDOP internal representation of the Earth system and not to provide state-of-the-art forecasts, only a selection of instruments has been used in training between 01/01/2013 and 01/01/2022; these are listed in Table~\ref{table:dop-observations} in the Appendix. This makes it possible to train GraphDOP on a single cluster node with 4 NVIDIA A100 graphics processing units (GPUs) and to run the forecast diagnostics described below on a single GPU. In our model configuration, there are approximately 50 million trainable parameters in the processor and 8 million parameters for the encoders and decoders.

\section{Interpretability in GraphDOP}
Although interpretability has escaped a precise and universal definition in the machine learning community, it can be defined in the context of GNNs for weather forecasting as the ability to understand the structure of the ML model, clarifying how different nodes (e.g. grid points), edges (e.g. physical connections), and features (e.g., temperature, pressure, wind speed) interact together \citep{kakkad2023,chen2024}. Interpretability is challenging in weather forecasting due to the complex nature of both the models and the data they are designed to handle. Nodes and edges interact through multiple layers of nonlinear transformations using latent variables rather than directly observable weather inputs. This section aims to illustrate how the structure of GraphDOP works to produce relevant and accurate weather forecasts. 

In machine learning, the Mean Squared Error (MSE) loss function can be defined as 
\begin{equation}
\mathcal{L}_{\text{MSE}}(\theta) = \frac{1}{N} \sum_{i=1}^{N} \left( y_i - f(\theta,x_i) \right)^2
\end{equation}
where $y_i$ is the target, $f(\theta,x_i)$ is the model prediction for parameters $\theta$ and input $x_i$, and $N$ is the number of samples. A standard Z-score standardization is applied to the dataset such that the mean is 0 and the standard deviation is 1. To find the optimal model parameters $\theta$ during the training of the ML model, backpropagation is typically used to compute the gradient of the model $f$ with respect to $\theta$. However, backpropagation can be extended further to compute the gradient of the model $f$ with respect to the input $x$ itself, by freezing the parameters $\theta$. This effectively builds the Jacobian matrix $J$ of the model in the normalized space where the i,j-th entry of $J$ contains the partial derivative of the i-th output with respect to the j-th input. A Jacobian-Vector Product (JVP) gives the sensitivity of the outputs to a change in the inputs, telling us in what direction the outputs of $f$ change if a perturbation to the inputs is made. It can be seen as a forward-mode in automatic differentiation and is equivalent to the Tangent Linear Model (TLM) used in DA to linearise the physical model while optimising the initial condition. Conversely, the Vector-Jacobian Product (VJP) provides the sensitivity of the inputs to changes in the outputs. In reverse-mode, the automatic differentiation tells us in what direction the inputs of $f$ change if a perturbation to the outputs is made. This way of computing derivatives is equivalent to the Adjoint Model (AD) which is also required in the 4D-Var minimisation \citep{Rabier00}. This adjoint model is a key component in computing Forecast Sensitivity to Observations (FSO) diagnostics which measure how sensitive a forecast is to individual observations \citep{Baker00,Ancell07,Zhu08}. It has also been used in the DA community to study the sensitivity of possible perturbations of initial conditions to forecast different physical variables \citep{Lopez01,Errico03,Mahfouf07}. Most deep learning frameworks (e.g. PyTorch, TensorFlow, JAX) utilise automatic differentiation, which enables the computation of gradients with respect to any differentiable input. As GraphDOP is based on PyTorch, the \texttt{torch.autograd.functional} module is used in this paper to compute (in reverse mode) the dot product between a perturbation vector $v$ and the Jacobian $J$. \citet{slivinski2024,tian2024,Solvik25} have already analysed the adjoint of different ML models to evaluate their possible use in a cycled DA system. To our knowledge, little work has been done so far on using them for an interpretability study.

\begin{figure}[htbp]
    \centering
    \includegraphics[width=1.0\textwidth]{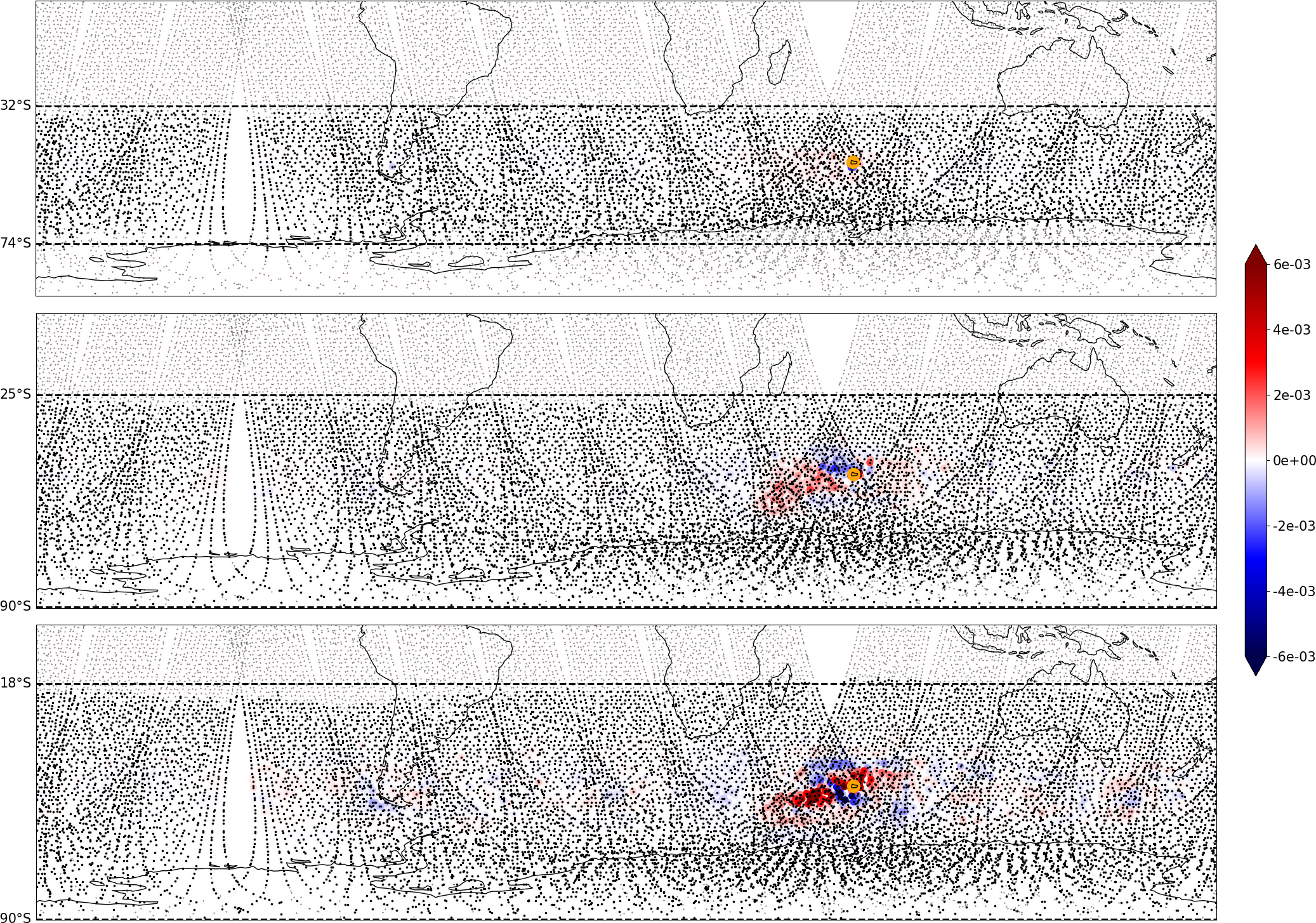}
    \caption{Sensitivity of input radiances (K) from NOAA-20 ATMS channel 6 to forecast 2m-temperature (C) in Kerguelen (orange dot) after 1500 (top), 9000 (middle) and 50000 (bottom) iterations in the training. This was computed for a forecast lead time of 12 hours on 01/01/2023.} 
    \label{fig:sensitivities_kerguelen}
\end{figure}

The reverse-mode automatic differentiation tool is used in this section to explain how the internal structure of GraphDOP unfolds during training when it is exposed to the different observation datasets. More specifically, we are interested in illustrating how DOP learns sensitivities between observations measuring different physical variables that are distributed in space and time. This work draws a parallel to the FSO diagnostics developed in traditional NWP to measure how sensitive a forecast is to individual observations. Figure~\ref{fig:sensitivities_kerguelen} shows the sensitivity of input radiances from NOAA-20-ATMS channel 6 to forecast 2-meter temperature in Kerguelen (orange dot) after 12 hours on 01/01/2023. These sensitivities are plotted after 1500 (top), 9000 (middle) and 50000 (bottom) training iterations (defined as the number of forward/backward passes with a batch). Kerguelen is one of the most remote islands with a weather station reporting in-situ measurements and is therefore an interesting location for a case study. ATMS channel 6 is sensitive to temperature with a peaking pressure level of 700hPa and is expected to play a role in the predictability of 2-meter temperature. The light grey dots in Figure~\ref{fig:sensitivities_kerguelen} show input radiances from NOAA-20-ATMS channel 6 with zero sensitivity to 2-meter temperature in Kerguelen, while the black dots show input radiances with a non-zero sensitivity. The GraphDOP attention window provides a built-in constraint on how processor nodes can influence each other as information can travel a certain distance within each processor layer \citep{lang2024}. The practical value of this inductive bias becomes clear when GraphDOP is trained on multiple observation datasets and learns the spatial and temporal input regions that are most relevant to the 12-hour weather forecast at any given point. This mechanism allows GraphDOP to explore possible correlations between microwave radiances and in-situ measurements. It exhibits variable latitude bands during the training from 32S-74S after 1500 iterations to 18S-90S after 50000 iterations. On the three different panels of Figure~\ref{fig:sensitivities_kerguelen}, the red (blue) circles show positive (negative) sensitivity that are significantly larger than zero. While these sensitivities are relatively small at the beginning of the training, they exhibit a clear pattern at the end, indicating that GraphDOP uses the information from radiances to predict the 2-meter temperature after 12 hours. The larger sensitivities located west of Kerguelen are consistent with the strong westerly winds that blow consistently from west to east and drive the weather around Antarctica. 

There is a fundamental link between the way GraphDOP learns its internal representation using attention, and DA systems using localization as a way to restrict updates to a local spatial neighbourhood. As models can be very high-dimensional and observations sparse and/or noisy, DA methods like the Ensemble Kalman Filter \citep{Evensen03} and ensemble-based 4D-Var \citep{isaksen2010eda} use localization to limit spurious long-range correlations in the covariance estimates. Additional effects of localization include reducing dimensionality (which aids implementation and parallelization), improving the effective use of information (e.g. in LETKF with limited degrees of freedom), and stabilizing matrix inversions through regularization \citep{Hunt07,Yan17}. This is typically done by applying a distance-based weighting function to the covariance matrix, often a compactly supported function like the one described in \citet{Gaspari99}. Such localization ensures that only observations within a certain spatial radius can significantly influence the model state update. In this context, localization can be seen as a hand-crafted attention mechanism that defines a weighting scheme for how much each observation affects each model variable. A difference is that localization in DA typically uses a fixed kernel (based on distance), whereas attention in ML learns an optimal weighting function from the training dataset.

\begin{figure}[htbp]
    \centering
    \includegraphics[width=0.85\textwidth]{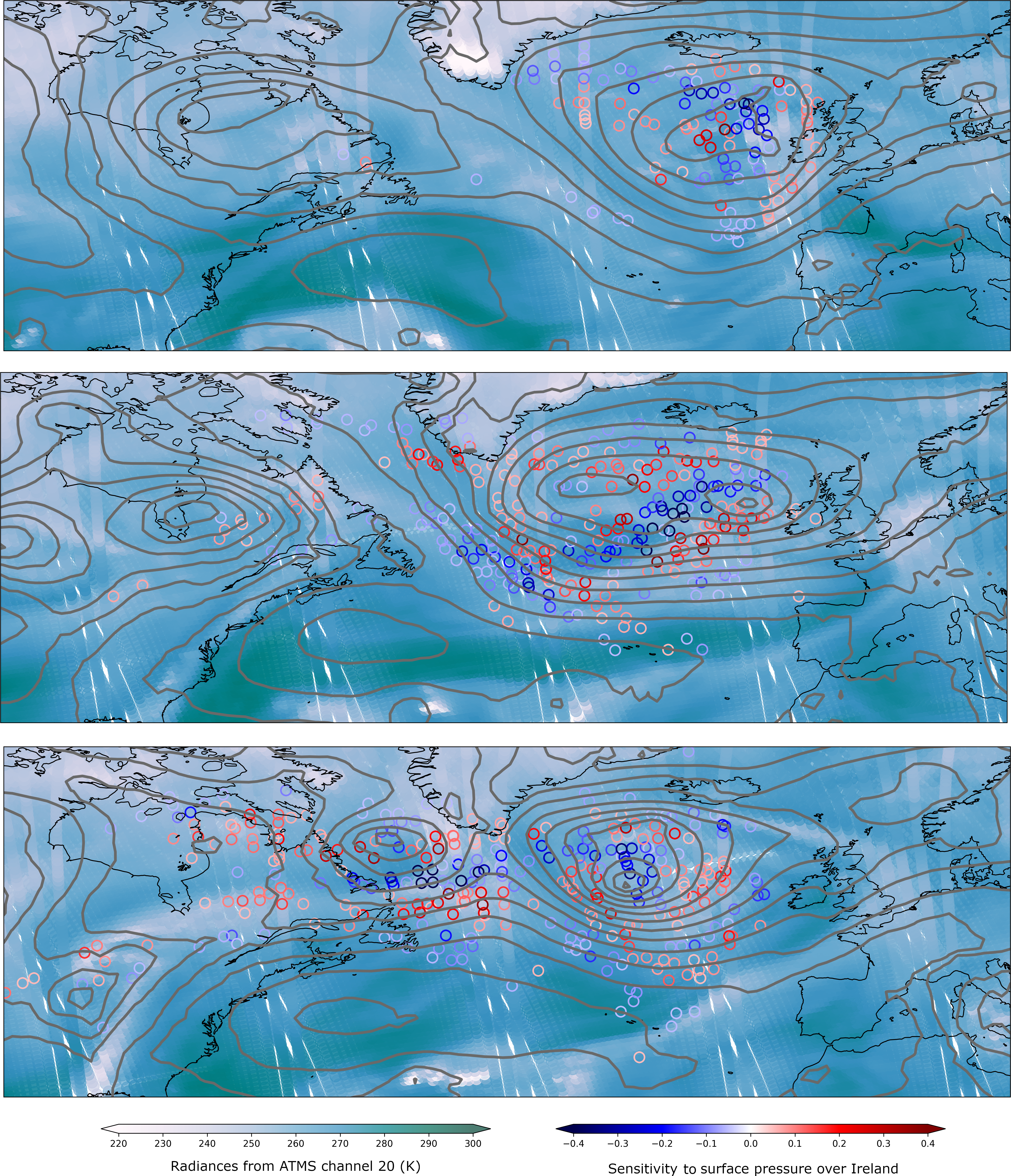}
    \caption{Sensitivity of input radiances (K) from NOAA-20-ATMS channel 20 (blue and red circles) to forecast surface pressure (hPa) in Ireland after 24h (top), 48h (middle) and 72h (bottom). The forecasts are respectively initialised on 07/11/2022, 06/11/2022 and 05/11/2022 to be all valid on 08/11/2022. The background shows all the input radiances from ATMS channel 20, and the contours show the synoptic situation at the forecast initial time.} 
    \label{fig:sensitivities_storm}
\end{figure}

The attention mechanism in GraphDOP is further illustrated through a case study from November 2022 when Hurricane Martin transitioned into an intense extratropical storm that moved further east across the North Atlantic and hit Ireland on the 8th of November. The red and blue circles in Figure~\ref{fig:sensitivities_storm} show the sensitivities of the input radiances from NOAA-20-ATMS channel 20 to forecast surface pressure in Ireland after 24h (top), 48h (middle), and 72h (bottom). The forecasts are respectively initialised on 07/11/2022, 06/11/2022, and 05/11/2022, and are all valid on 08/11/2022. The background shows all the input radiances from ATMS channel 20, which is sensitive to water vapour in the mid-troposphere, and the contours show the corresponding synoptic situation at the forecast initial time.

\textcolor{black}{Water vapour acts as a natural tracer in the atmosphere as its movement is linked to atmospheric circulation, and it can be used to follow the movement of air masses. It is striking to see how GraphDOP uses the water vapour information to track the evolution of the different low-pressure systems.} For the 24h forecast (top panel), the most important information used to predict the surface pressure over Ireland comes slightly west where the extratropical storm is currently located. The attention area is moving further west for the 48h forecast (middle panel) and 72h forecast (bottom panel), where GraphDOP becomes more sensitive to radiances located around the different low-pressure systems.  This case study demonstrates how GraphDOP learned from observational data to use humidity information to track pressure systems and to shift its attention further east with lead time. It shows that the nodes in the GraphDOP processor exchange information with their neighbours to move the latent space forward in time. \textcolor{black}{It is similar in concept to the physical dynamics core of a traditional numerical weather model and a 4D-Var system where Navier–Stokes equations describe the motion of fluids and humidity radiances drive advection wind tracing.} \textcolor{black}{These are arguably speculations and more work will be needed to gain a better understanding.}

\section{Explainability in GraphDOP}
While interpretability refers to the degree to which a human can understand the internal mechanics of an ML model, explainability refers to the extent to which humans can comprehend and rationalise the predictions of an ML model. This concept of explainability has been explored in the context of weather forecasting \citep{Marcin2023,yang2024} where the ML community developed many tools to estimate the feature importance, such as SHapley Additive exPlanations (SHAP, \citet{lundberg2017unified}) and Local Interpretable Model-Agnostic Explanations (LIME, \citet{ribeiro2016should}). In parallel, the NWP community faced similar challenges when evaluating the importance of different instruments in producing valuable weather forecasts. A traditional way of estimating data impact in a forecasting system is to perform OSEs. This involves removing one or more datasets from the full observing system over a long assimilation period. One then compares forecast skill against that of a control experiment, which assimilates the available observations from the global network \citep{Bouttier01}. To complement the results from OSEs, many NWP centres developed their own Forecast Sensitivity-based Observation Impact (FSOI) diagnostic tool to estimate how each observation changes the forecast error \citep{Baker00,Cardinali09,Lorenc14}. This section aims to apply the FSOI method in the context of the GraphDOP model to study its explainability and assess whether it differs from a classic NWP model. It complements the work in \citet{Lean25} and \citet{Boucher25} where a series of experiments is conducted showing that GraphDOP develops an internal representation of the Earth system.

The use of adjoint-based estimates of observation sensitivities to calculate the impact of observations on forecast errors was introduced by \citet{Baker00}. The computation of the FSOI value is based on the definition of a quadratic measure of forecast error $e$ with respect to an initial condition $x$ and for a given lead time $h$
\begin{equation}\label{error}
e(x)=\left({\cal M}_{0\rightarrow h}(x)-x^{ref}\right)^TC\left({\cal M}_{0\rightarrow h}(x)-x^{ref}\right)
\end{equation}
where ${\cal M}_{0\rightarrow h}$ is the forecasting model, $x^{ref}$ is the reference state valid at the same time as the forecast, and $C$ is a symmetric (more typically diagonal) weighting matrix. The specification of $C$ is explained below when the different experiments are discussed. Although the forecast error $e$ is a quadratic expression in ${\cal M}_{0\rightarrow h}$, it becomes a higher-order expression when it is expressed in terms of the initial condition $x$. The change in the forecast error when observations are assimilated can be estimated by computing a Taylor expansion to the third order around the background state $x_b$ and evaluated at the analysis state $x_a$ 
\begin{equation}\label{FSOI}
e(x_a)-e(x_b)=\left(x_a-x_b\right)^T\left(M^T_bC\left({\cal M}_{0\rightarrow h}(x_b)-x^{ref}\right)+M^T_aC({\cal M}_{0\rightarrow h}(x_a)-x^{ref})\right)
\end{equation}
where $M^T_b$ is the adjoint of the model linearised around $x_b$ and $M^T_a$ is the adjoint of the model linearised around $x_a$. They are both computed using the VJP product implemented in PyTorch. A comprehensive derivation of Equation~\ref{FSOI} is given in \citet{Errico07}.

In a standard assimilation system, the change in the forecast error is linked to the assimilated observations $y$ by replacing $(x_a-x_b)$ with the expression of the analysis increment $K(y-{\cal H}x_b)$ where $K$ is the Kalman gain matrix and ${\cal H}$ is the observation operator. This step is not necessary with GraphDOP as it already operates in the observation space where $x_b$ is a 12h forecast initialised from the previous input window and $x_a$ is the observations from the current input window. The FSOI value for the i-th observation in the current input window is then given by the i-th component of the dot product in Equation~\ref{FSOI}. A negative value means that the i-th observation reduces the forecast error, while a positive value means that the observation has a detrimental impact and increases the forecast error. There are several possibilities on how to specify the forecast error metric defined in Equation~\ref{error}. By using different weighting matrices $C$, the difference between the forecast and the truth can be computed for a specific meteorological variable or for a specific geographical area. It is also possible to look at the impact of observations for different lead times by specifying a different integration length $h$ for the forecasting model.

\begin{figure}[htbp]
    \centering
    \includegraphics[width=\textwidth]{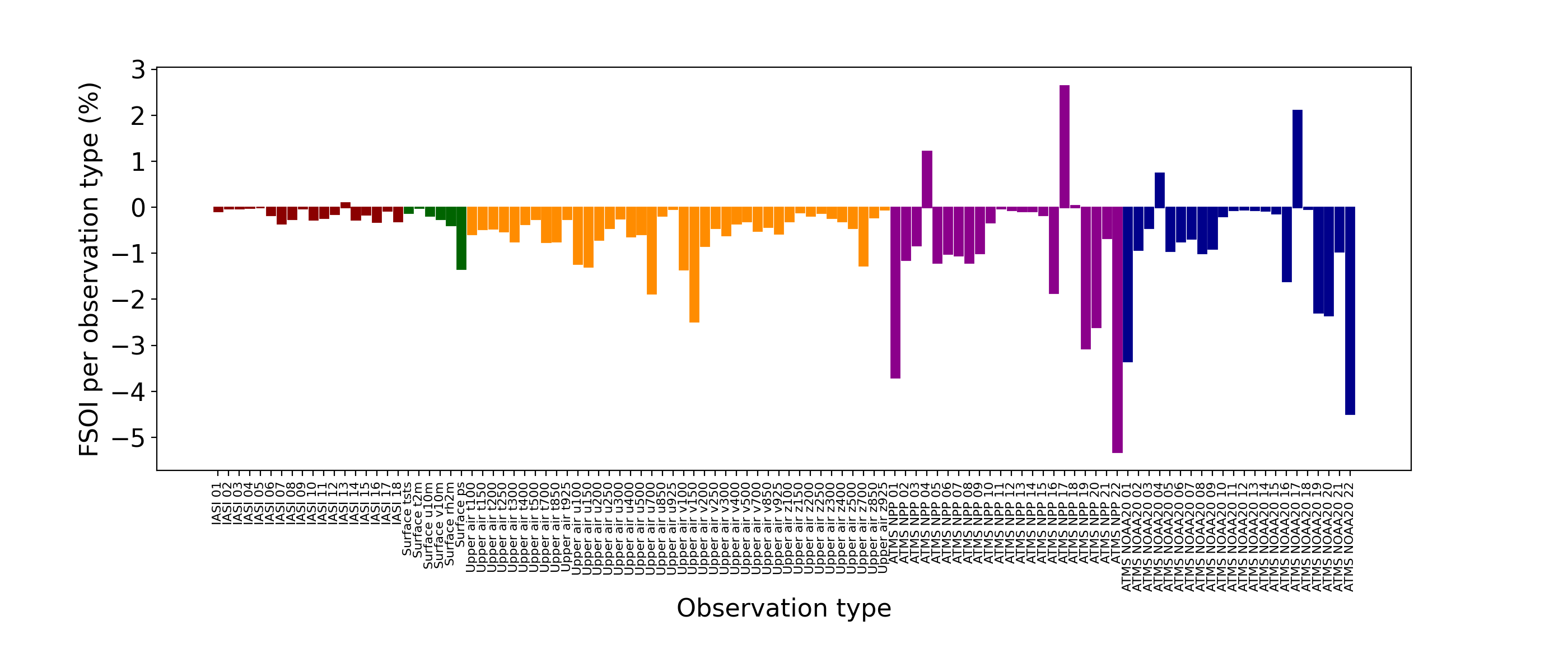}
    \caption{Relative contribution of each data type in the global forecast error after 12 hours. Negative (positive) values correspond to a decrease (increase) in the forecast error. Statistics have been averaged between 01/01/2023 and 01/03/2023.} 
    \label{fig:fsoi_global}
\end{figure}

\textcolor{black}{A standard diagnostic is to compute the mean FSOI value for the different data types (e.g. satellite channels, radiosonde pressure levels and surface variables) and to plot the relative contribution of each to the change in the global forecast error across all observations. To take into account the difference in the number of observations per data type, the mean FSOI value is computed for each of them. In this work, the global forecast error is not weighted according to the number of observations (i.e. $C$ weighting matrix is set to the identity). This means that the forecast error value is dominated by data types with larger numbers of observations.}
Figure~\ref{fig:fsoi_global} shows the relative contribution of each data type after 12 hours, where FSOI statistics are computed over all observations and all locations and averaged between 01/01/2023 and 01/03/2023. It is promising for the GraphDOP forecasting system to have most data types working together to reduce the forecast error. Microwave channels from ATMS (purple and blue bars) sensitive to the surface (channel 1) and to humidity (channel 22) have particularly large FSOI values. \textcolor{black}{Conversely, channels 4 and 17 have positive FSOI values, meaning that these types increase the forecast error. These two channels contain a mixture of surface and atmospheric sensitivity which might be harder for GraphDOP to interpret. Channel 4 is a surface channel with a lot of tropospheric temperature sensitivity, while channel 17 is a "dirty" surface channel seeing mostly surface in dry conditions and water vapour in the tropics, with sensitivity to snow/graupel and liquid clouds.}

\begin{figure}[htbp]
    \centering
    \includegraphics[width=1.0\textwidth]{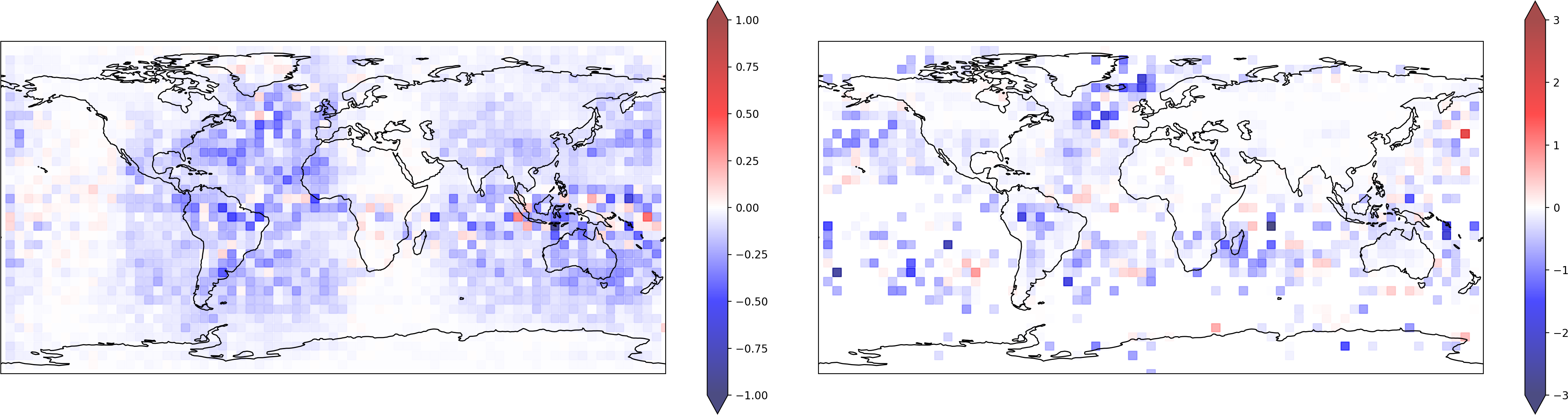}
    \caption{Relative contribution of ATMS channel 22 (left) and surface pressure (right) in the global forecast error after 12 hours. Negative (positive) values correspond to a decrease (increase) in the forecast error. Statistics have been averaged between 01/01/2023 and 01/03/2023.} 
    \label{fig:fsoi_map}
\end{figure}

\textcolor{black}{Since an FSOI value is computed for each observation in the input window, geographical maps can be produced to study the spatial significance of measurements from a given data type. We remind the reader that the input window denotes the time period where input observations are used to run GraphDOP in inference mode (i.e., 12 hours in this work) and is not related to the sliding attention window used in the processor and discussed earlier in the paper.}
The left panel of Figure~\ref{fig:fsoi_map} shows a map of the relative contribution of ATMS channel 22, which is one of the largest contributors to reducing the forecast error. Only grids with more than 300 measurements are plotted to get significant results \citep{Lorenc14}. The two areas that make the largest contributions are those which are valid near the end of the input window. \textcolor{black}{We may speculate that such behaviour can be explained as the observations valid at the end of the input window contain the latest snapshot of the atmospheric state.} \textcolor{black}{In 4D-Var, a similar larger sensitivity to observations valid near the end of the assimilation window is also present and can be explained by the same argument. The larger sensitivity also comes in 4D-Var from the forecast model that can evolve numerous atmospheric variables over the assimilation window to fit the data, doing for example wind tracing \citep{Mcnally19}.} The right panel of Figure~\ref{fig:fsoi_map} shows a similar map for surface pressure observations. While most measurements are made over Europe and the USA, observations in remote places, such as over oceans, have a larger impact. This behaviour has also been noticed in 4D-Var when assimilating extra observations from newly available platforms. 

The FSOI diagnostics presented in Figure~\ref{fig:fsoi_global} and Figure~\ref{fig:fsoi_map} provide valuable insights into the information content of the different data types. In the IFS system, a careful selection of the different instruments and channels entering the system is made. This is because the physical processes captured by observations need to be accurately represented in the IFS to make their assimilation possible. Otherwise, 4D-Var cannot correctly address the mismatch between the spatial and temporal scales represented by observations and the model's representation of the atmosphere. Such a requirement is not necessary in GraphDOP as any observational dataset can be \textcolor{black}{technically} added during the training. In this context, the FSOI diagnostic could help build a high-quality training dataset, avoiding the introduction of confusing information.

\begin{figure}[htbp]
    \centering
    \includegraphics[width=1.0\textwidth]{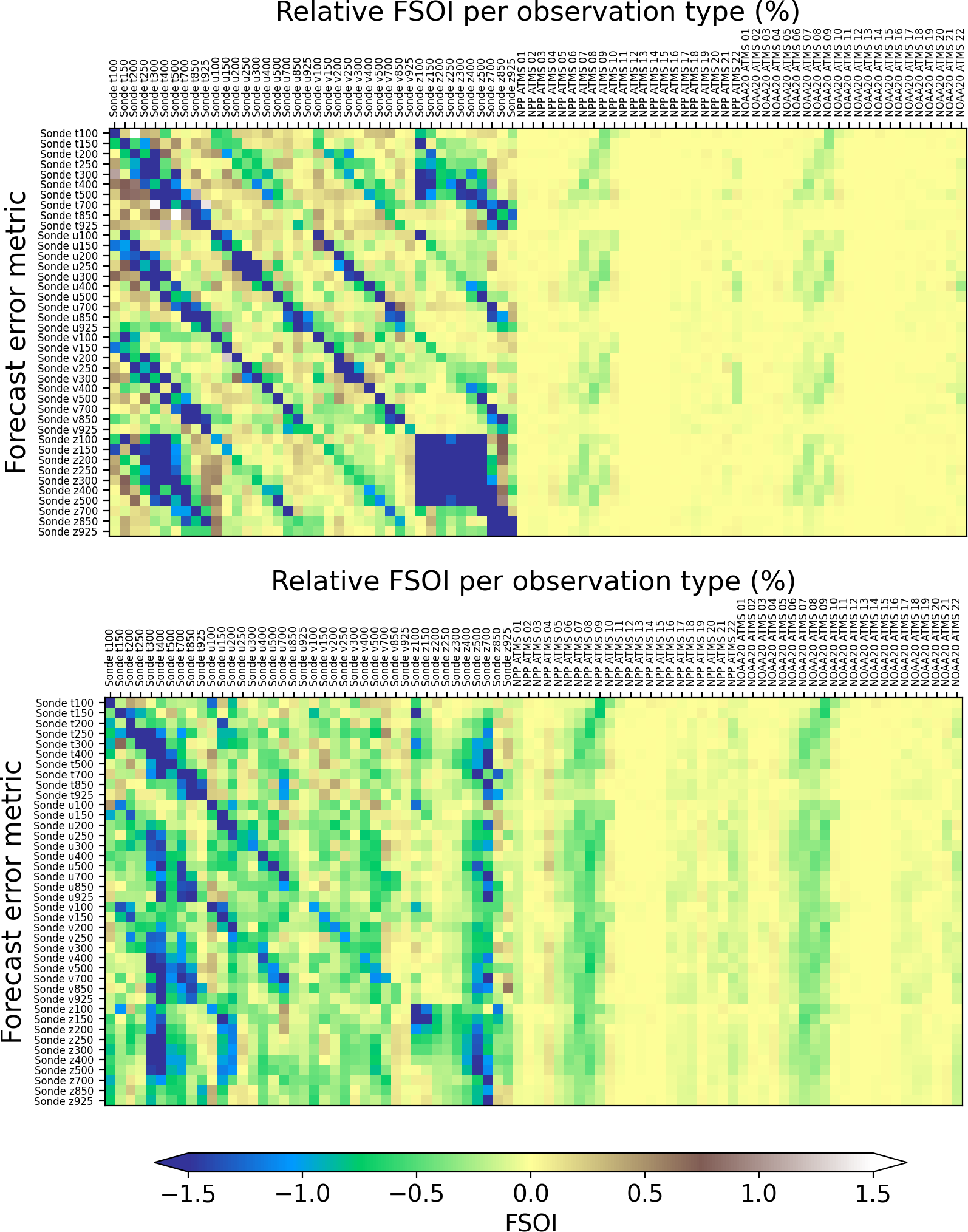}
    \caption{Relative contribution of each data type (columns) for different forecast error metrics (rows) after 12 hours (top panel) and 36 hours (bottom panel). Statistics have been averaged between 01/01/2023 and 01/03/2023.} 
    \label{fig:fsoi_matrix}
\end{figure}

To further discuss the explainability of GraphDOP, we look at the impact of the different data types to predict the main physical variables (i.e. temperature, zonal wind, meridional wind and geopotential) for the different pressure levels (i.e. 100, 150, 200, 250, 300, 400, 500, 700, 850, 925 hPa). This can be done by defining a set of forecast error metrics where the weighting matrix ($C$ in Equation~\ref{error}) selects only measurements from radiosondes for one physical variable and one pressure level at a time. By computing the FSOI values for all observations available in the input window using each of these forecast error metrics, we can study the origin of GraphDOP predictability and gain a better insight into how GraphDOP builds its internal representation of the Earth system. Figure~\ref{fig:fsoi_matrix} summarises the results where each row represents a different forecast error metric and each column represents a different data type. The top panel shows the relative FSOI after 12 hours averaged between 01/01/2023 and 01/03/2023. The main dark blue diagonal shows that GraphDOP reduces the forecast error of a given parameter at a given pressure level by relying primarily on the information contained in the corresponding radiosonde measurements. This demonstrates that GraphDOP accurately captures the vertical structure present in the radiosonde input dataset. The blue off-diagonal terms show some cross correlations between the physical fields, for example, the importance of temperature in reducing wind errors. The different forecast errors are also sensitive to the ATMS tropospheric temperature channels (6 to 10). It illustrates that GraphDOP successfully extracts information from radiances to predict temperature, wind, and geopotential. The reliance on microwave information is even stronger when the lead time is increased to 36 hours (see bottom panel of Figure~\ref{fig:fsoi_matrix}). This plot provides a clear evidence that understanding how these radiances evolve in space and time is crucial for GraphDOP to produce competitive forecast skill scores. Note that the FSOI values for the different infrared channels are not plotted here to improve readability.

\begin{figure}[htbp]
    \centering
    \includegraphics[width=1.0\textwidth]{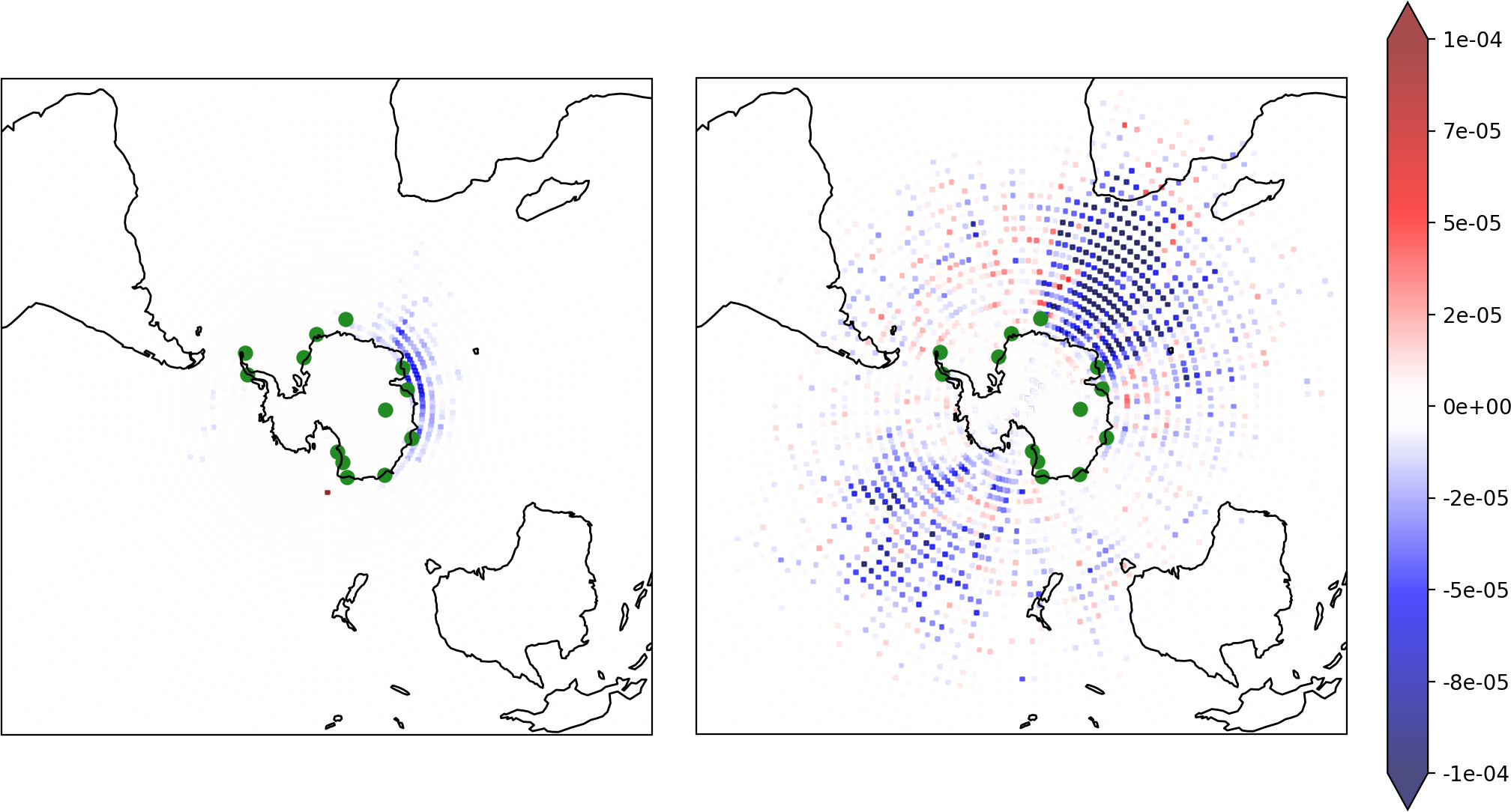}
    \caption{Relative contribution of IASI wavenumber 756 in the change of 850hPa temperature forecast error at the radiosonde location (green circles) after 1 day (left) and 4 days (right). Statistics have been averaged between 01/01/2023 and 01/03/2023.} 
    \label{fig:fsoi_spole}
\end{figure}

We now describe a case study performed over the South Pole to further highlight the importance of satellite observations to predict temperature fields. This is a relevant area with a sparse in-situ network that contains a limited number of radiosonde stations around the Antarctic coast (green circles in Figure~\ref{fig:fsoi_spole}). \textcolor{black}{The region is, however, well observed by polar orbiting satellites (METOP-B IASI, NOAA-20 ATMS and NPP ATMS in our experiments). We showed in Figure~\ref{fig:fsoi_global} that microwave sounders have the largest contribution to reduce the global forecast error across all observations. Results are different when the forecast error is computed only for temperature measurements from the green radiosonde stations at 850hPa after 1 day. The FSOI diagnostic shows that the relative impact of radiosondes observations (e.g. -3\% for temperature at 850hPa) is slightly larger than the impact of different satellite radiances (e.g. -2\% for IASI wavenumber 756 radiances or -1\% for ATMS channel 1 radiances). It is interesting to see in this case the larger importance of IASI wavenumber 756, which is a temperature/surface channel with a weighting function that peaks at 930 hPa and ATMS channel 1, which is sensitive to water vapour and clouds in the boundary layer. It is sensible to see these two channels having a role in the prediction of temperature at 850hPa}. The importance of IASI wavenumber 756 is further explored by plotting a geographical map of the FSOI values (see left panel of Figure~\ref{fig:fsoi_spole}). It shows that the impact of the satellite channel stays located extremely close to the coast where the radiosonde stations are located. The relative importance of the different data types changes drastically when the forecast lead time is increased from 1 day to 4 days. The impact of radiosonde measurements decreases (e.g. 0.5\% for temperature at 850hPa) while the importance of satellite channels increases (e.g. -3\% for IASI wavenumber 756 or -2\% for ATMS channel 1). It is illustrated on the right panel of Figure~\ref{fig:fsoi_spole} where the relative contribution of IASI wavenumber 756 is plotted after 4 days. For both lead times, there is only a very small contribution from these radiances over land as it is always covered with ice and presents the same brightness temperature signature. This case study illustrates how GraphDOP relies significantly on satellite channels to forecast the weather in remote location where few in-situ observations are available. This is especially true for longer lead times where GraphDOP needs to extract the information from satellite radiances properly to obtain the correct global synoptic situation. 
\textcolor{black}{Further investigations are required to better understand why some channels (e.g. IASI wavenumber 756) have a relative small impact for the global forecast error accross all variables and a larger one when computed over a selection of radiosondes.}

\begin{figure}[htbp]
    \centering
    \includegraphics[width=\textwidth]{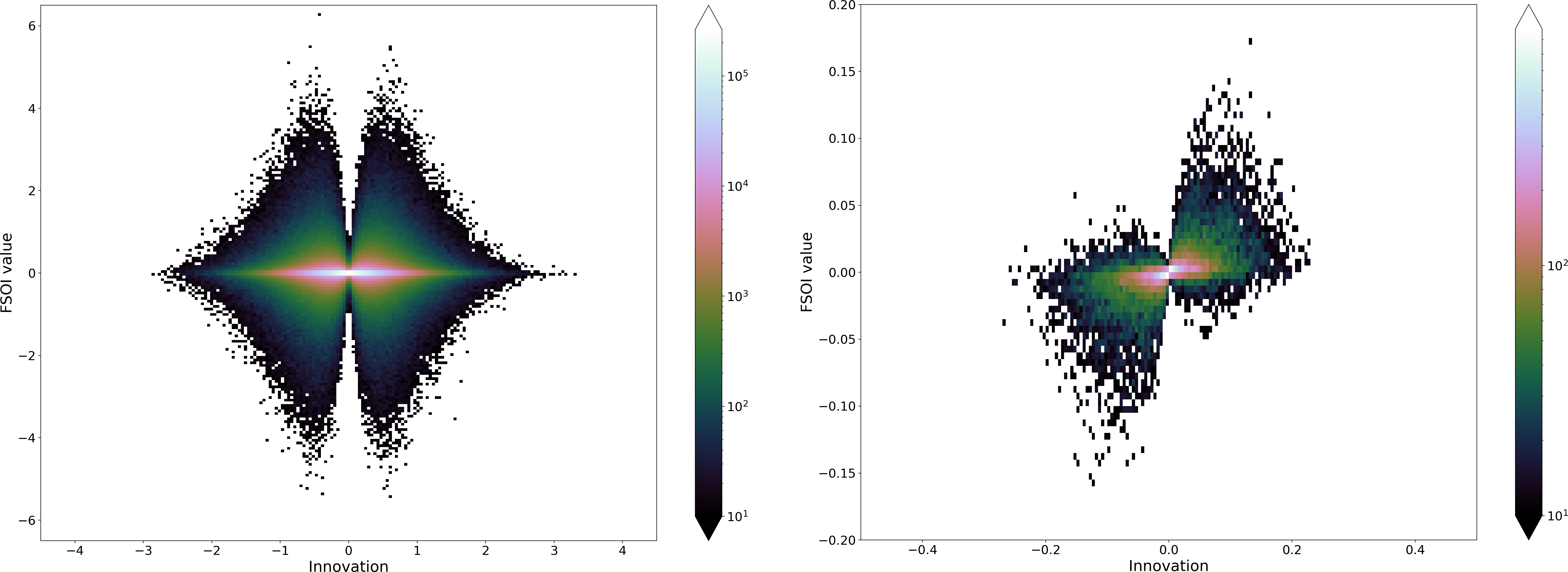}
    \caption{Scatter plot of normalised innovations against FSOI values for the global forecast error of 10m zonal wind after 12 hours (left) and for the local forecast error of 2-meter temperature over Northern Canada after 12 hours (right). Statistics have been averaged between 01/01/2023 and 01/03/2023.} 
    \label{fig:fsoi_scatter}
\end{figure}

Scatter plots are traditionally used to study the relationship between innovations and FSOI values. Innovations are usually defined as the difference between the model equivalents and the observations. In the GraphDOP context, it is given by the difference between a 12h forecast initialised from the previous input window and the observations from the current input window. \citet{Lorenc14} showed that observations further away from the model (i.e. with a larger innovation) have a larger impact in a DA system, characterised by statistically larger FSOI values (positive or negative). It is illustrated for GraphDOP on the left panel of Figure~\ref{fig:fsoi_scatter}, showing the normalised innovations from the 10m zonal wind measurements and their FSOI impacts on the global forecast error of 10m zonal wind after 12 hours. This butterfly pattern is observed for most instruments/channels. To some extend, it is matching the cone pattern showed in Figure 18 of \citet{Lorenc14}, although the FSOI values in GraphDOP are decreasing for the largest innovations located at the left and right tails of the distribution. The right panel of Figure~\ref{fig:fsoi_scatter} shows the normalised innovations from 2-meter temperature over Northern Canada and their corresponding FSOI impacts on the forecast error of 2-meter temperature over the same region after 12 hours. The scatter plot displays an asymmetric shape that can be explained by the small cold bias in the GraphDOP model, which is a few tenths of a Kelvin over Northern Canada. In such a situation, an observation cooler than the model (i.e. with a positive innovation) will reinforce the cold model bias and is statistically more likely to increase the forecast error (i.e. have a positive FSOI). Conversely, an observation warmer than the model (i.e. with a negative innovation) will counteract the cold model bias and is statistically more likely to reduce the forecast error (i.e. have a negative FSOI). One must be cautious when interpreting FSOI applied to a real system that could contain model biases \citep{Liu08,Necker18,Prive21}.

\section{Conclusions and future works}

The Data Assimilation community has been developing various sensitivity-based diagnostics to understand and evaluate the importance of the observing system in accurately forecasting the weather. \textcolor{black}{The Forecast Sensitivity-based Observation Impact (FSOI) tool has been used for decades in many NWP centres to measure the impact of each observation on the forecast error. This paper examines how these diagnostics can be applied to GraphDOP to enhance its interpretability and explainability.} From the computation of the GraphDOP Jacobian, results demonstrate how GraphDOP captures real, physically meaningful processes, such as the movement of storm systems. We also provided FSOI-based examples where GraphDOP constructs a coherent internal representation of the Earth system by blending information coming from heterogeneous input data, specifically conventional observations and satellite radiances.  

The results presented in this paper are based on a version of GraphDOP trained specifically for a limited number of observation types. The primarily goal was to reduce the computational time and run the adjoint-based tools on a single GPU. A natural extension of this work is to replicate it with the current state-of-the-art model and evaluate the importance of the different observation types. This could help in understanding whether the model relies too much or too little on a specific observation type and inform the creation of new training datasets by identifying gaps. 

Interpretability and explainability play a crucial role in diagnosing errors and improving models. When a forecast is incorrect, it is crucial to have diagnostic tools to dissect the ML model and understand which steps in the process influenced the result. Such a feedback is essential for refining graph structures, tuning feature representations, identifying weaknesses in the training datasets, and ultimately improving prediction quality. We believe that the tools presented in the paper can be applied in the future to benchmark the quality of different AI-DOP models, to ensure that they provide the correct forecast for the right reason.

\textcolor{black}{Another prospect of this work is to implement an online observation Quality Control (QC) mechanism based on the impact measured by the FSOI diagnostic. Such an idea has been explored in standard assimilation and forecasting system with some reasonable degree of success \citep{Daisuke17,Tse20}. The FSOI diagnostics could also be complemented by OSEs experiments, withdrawing satellite channels or conventional stations (from the training dataset or the inference input window) that consistently exhibit positive FSOI values. These strands of work and their impact on forecast skill scores have not been explored yet.}

\section{Acknowledgements}
Many people have contributed in various ways to AI-DOP at ECMWF. We would like to thank Mat Chantry, Peter Dueben, Christian Lessig, Niels Bormann, Sean Healy, Patricia de Rosnay, Massimo Bonavita, Florian Pinault, Mario Santa Cruz, Chris Burrows, Alan Geer, Bruce Ingleby, Philippe Lopez and David Duncan.
\bibliographystyle{unsrtnat}
\bibliography{dop}

\begin{thebibliography}{53}
\providecommand{\natexlab}[1]{#1}
\providecommand{\url}[1]{\texttt{#1}}
\expandafter\ifx\csname urlstyle\endcsname\relax
  \providecommand{\doi}[1]{doi: #1}\else
  \providecommand{\doi}{doi: \begingroup \urlstyle{rm}\Url}\fi

\bibitem[Venuti et~al.(2025)Venuti, Rabier, Andersson, Modigliani, English, Kitchen, Berrone, Pinsault, Debrux, and Jarraud]{Venuti25}
Fabio Venuti, Florence Rabier, Erik Andersson, Umberto Modigliani, Stephen English, Christine Kitchen, Matthieu Berrone, Dorian Pinsault, Julie Debrux, and Michel Jarraud.
\newblock {ECMWF}’s societal impact through service provision, partnerships and collaborations.
\newblock \emph{Journal of the European Meteorological Society}, 2:\penalty0 1--11, 2025.
\newblock \doi{10.1016/j.jemets.2025.100013}.

\bibitem[Dahoui and Sahin(2024)]{Dahoui24}
Mohamed Dahoui and Cihan Sahin.
\newblock On-demand web plotting of observation monitoring statistics.
\newblock \emph{{ECMWF} Newsletter No. 178}, 2024.
\newblock \doi{10.21957/md3v5hk9ge}.

\bibitem[Bouttier and Kelly(2001)]{Bouttier01}
F.~Bouttier and G.~Kelly.
\newblock Observing-system experiments in the {ECMWF} {4D-Var} data assimilation system.
\newblock \emph{Quarterly Journal of the Royal Meteorological Society}, 127\penalty0 (574):\penalty0 1469--1488, 2001.
\newblock \doi{10.1002/qj.49712757419}.

\bibitem[Bormann et~al.(2019)Bormann, Lawrence, and Farnan]{Bormann19}
Niels Bormann, Heather Lawrence, and Jacky Farnan.
\newblock Global observing system experiments in the {ECMWF} assimilation system.
\newblock \emph{ECMWF Technical Memoranda No. 839}, 2019.
\newblock \doi{10.21957/sr184iyz}.

\bibitem[Bonavita et~al.(2012)Bonavita, Isaksen, and Hólm]{Bonavita12}
Massimo Bonavita, Lars Isaksen, and Elías Hólm.
\newblock On the use of {EDA} background error variances in the {ECMWF} {4D-Var}.
\newblock \emph{Quarterly Journal of the Royal Meteorological Society}, 138\penalty0 (667):\penalty0 1540--1559, 2012.
\newblock \doi{10.1002/qj.1899}.

\bibitem[Rodwell et~al.(2016)Rodwell, Lang, Ingleby, Bormann, Hólm, Rabier, Richardson, and Yamaguchi]{Rodwell16}
M.~J. Rodwell, S.~T.~K. Lang, N.~B. Ingleby, N.~Bormann, E.~Hólm, F.~Rabier, D.~S. Richardson, and M.~Yamaguchi.
\newblock Reliability in ensemble data assimilation.
\newblock \emph{Quarterly Journal of the Royal Meteorological Society}, 142\penalty0 (694):\penalty0 443--454, 2016.
\newblock \doi{10.1002/qj.2663}.

\bibitem[Buizza(2019)]{Buizza19}
Roberto Buizza.
\newblock Introduction to the special issue on “25 years of ensemble forecasting”.
\newblock \emph{Quarterly Journal of the Royal Meteorological Society}, 145\penalty0 (S1):\penalty0 1--11, 2019.
\newblock \doi{10.1002/qj.3370}.

\bibitem[Rabier et~al.(2000)Rabier, Järvinen, Klinker, Mahfouf, and Simmons]{Rabier00}
F.~Rabier, H.~Järvinen, E.~Klinker, J.-F. Mahfouf, and A.~Simmons.
\newblock The {ECMWF} operational implementation of four-dimensional variational assimilation. {I}: Experimental results with simplified physics.
\newblock \emph{Quarterly Journal of the Royal Meteorological Society}, 126\penalty0 (564):\penalty0 1143--1170, 2000.
\newblock \doi{10.1002/qj.49712656415}.

\bibitem[Baker and Daley(2000)]{Baker00}
Nancy~L. Baker and Roger Daley.
\newblock Observation and background adjoint sensitivity in the adaptive observation-targeting problem.
\newblock \emph{Quarterly Journal of the Royal Meteorological Society}, 126\penalty0 (565):\penalty0 1431--1454, 2000.
\newblock \doi{https://doi.org/10.1002/qj.49712656511}.

\bibitem[Langland and Baker(2004)]{Langland04}
Rolf Langland and Nancy Baker.
\newblock Estimation of observation impact using the {NRL} atmospheric variational data assimilation adjoint system.
\newblock \emph{Tellus A}, 56\penalty0 (3):\penalty0 189--201, 2004.
\newblock \doi{10.3402/tellusa.v56i3.14413}.

\bibitem[Cardinali(2009)]{Cardinali09}
Carla Cardinali.
\newblock Monitoring the observation impact on the short-range forecast.
\newblock \emph{Quarterly Journal of the Royal Meteorological Society}, 135\penalty0 (638):\penalty0 239--250, 2009.
\newblock \doi{10.1002/qj.366}.

\bibitem[Keisler(2022)]{keisler2022}
Ryan Keisler.
\newblock Forecasting global weather with {G}raph {N}eural {N}etworks.
\newblock \emph{arXiv preprint arXiv:2202.07575}, 2022.
\newblock \doi{10.48550/arXiv.2202.07575}.

\bibitem[Lam et~al.(2023)Lam, Sanchez-Gonzalez, Willson, Wirnsberger, Fortunato, Alet, Ravuri, Ewalds, Eaton-Rosen, Hu, Merose, Hoyer, Holland, Vinyals, Stott, Pritzel, Mohamed, and Battaglia]{lam2023}
Remi Lam, Alvaro Sanchez-Gonzalez, Matthew Willson, Peter Wirnsberger, Meire Fortunato, Ferran Alet, Suman Ravuri, Timo Ewalds, Zach Eaton-Rosen, Weihua Hu, Alexander Merose, Stephan Hoyer, George Holland, Oriol Vinyals, Jacklynn Stott, Alexander Pritzel, Shakir Mohamed, and Peter Battaglia.
\newblock Learning skillful medium-range global weather forecasting.
\newblock \emph{Science}, 382\penalty0 (6677):\penalty0 1416--1421, 2023.
\newblock \doi{10.1126/science.adi2336}.
\newblock URL \url{https://www.science.org/doi/10.1126/science.adi2336}.

\bibitem[Lang et~al.(2024{\natexlab{a}})Lang, Alexe, Chantry, Dramsch, Pinault, Raoult, Clare, Lessig, Maier-Gerber, Magnusson, Bouall{\`e}gue, Nemesio, Dueben, Brown, Pappenberger, and Rabier]{lang2024}
Simon Lang, Mihai Alexe, Matthew Chantry, Jesper Dramsch, Florian Pinault, Baudouin Raoult, Mariana C.~A. Clare, Christian Lessig, Michael Maier-Gerber, Linus Magnusson, Zied~Ben Bouall{\`e}gue, Ana~Prieto Nemesio, Peter~D. Dueben, Andrew Brown, Florian Pappenberger, and Florence Rabier.
\newblock {AIFS} -- {ECMWF}'s data-driven forecasting system, 2024{\natexlab{a}}.
\newblock URL \url{https://arxiv.org/abs/2406.01465}.

\bibitem[Price et~al.(2025)Price, Sanchez-Gonzalez, Alet, Andersson, El{-}Kadi, Masters, Ewalds, Stott, Mohamed, Battaglia, Lam, and Willson]{Price2025GenCast}
Ilan Price, Alvaro Sanchez-Gonzalez, Ferran Alet, Tom~R. Andersson, Andrew El{-}Kadi, Dominic Masters, Timo Ewalds, Jacklynn Stott, Shakir Mohamed, Peter Battaglia, Remi Lam, and Matthew Willson.
\newblock Probabilistic weather forecasting with machine learning.
\newblock \emph{Nature}, 637\penalty0 (8044):\penalty0 84--90, January 2025.
\newblock \doi{10.1038/s41586-024-08252-9}.
\newblock URL \url{https://doi.org/10.1038/s41586-024-08252-9}.

\bibitem[Lang et~al.(2024{\natexlab{b}})Lang, Alexe, Clare, Roberts, Adewoyin, Bouall\`{e}gue, Chantry, Dramsch, Dueben, Hahner, Maciel, Prieto{-}Nemesio, O'Brien, Pinault, Polster, Raoult, Tietsche, and Leutbecher]{Lang2024AIFSCRPS}
Simon Lang, Mihai Alexe, Mariana C.~A. Clare, Christopher Roberts, Rilwan Adewoyin, Zied~Ben Bouall\`{e}gue, Matthew Chantry, Jesper Dramsch, Peter~D. Dueben, Sara Hahner, Pedro Maciel, Ana Prieto{-}Nemesio, Cathal O'Brien, Florian Pinault, Jan Polster, Baudouin Raoult, Steffen Tietsche, and Martin Leutbecher.
\newblock {AIFS{-}CRPS}: Ensemble forecasting using a model trained with a loss function based on the continuous ranked probability score, 2024{\natexlab{b}}.
\newblock URL \url{https://arxiv.org/abs/2412.15832}.

\bibitem[Hersbach et~al.(2020)Hersbach, Bell, Berrisford, Hirahara, Hor\'anyi, noz\textendash Sabater, Nicolas, Peubey, Radu, Schepers, Simmons, Soci, Abdalla, Abellan, Balsamo, Bechtold, Biavati, Bidlot, Bonavita, Giovanna~De\, Dahlgren, Dee, Diamantakis, Dragani, Flemming, Forbes, Fuentes, Geer, Haimberger, Healy, Robin\, H\'olm, Janiskov\'a, Stephen\, Laloyaux, Lopez, Lupu, Radnoti, Richardson, Trevisan, Tr\'emolet, Vamborg, Villaume, and Th\'epaut]{hersbach2020era5}
Hans Hersbach, Bill Bell, Paul Berrisford, Shoji Hirahara, Andr\'as Hor\'anyi, Joaquim~Mu\ noz\textendash Sabater, J\'er\^ome Nicolas, Caroline Peubey, Raluca Radu, Dian Schepers, Adrian Simmons, Cornel Soci, Saleh Abdalla, Xavier Abellan, Gianpaolo Balsamo, Peter Bechtold, Gianfranco Biavati, Jean\textendash~Raymond Bidlot, Massimo Bonavita, Chiara Giovanna~De\, Per Dahlgren, Dick Dee, Marios Diamantakis, Rossana Dragani, Johannes Flemming, Richard Forbes, Manuel Fuentes, Alan Geer, Leo Haimberger, Sean Healy, J.~Hogan Robin\, Esben H\'olm, Marta Janiskov\'a, E.~Keeley Stephen\, P.\, Patrick Laloyaux, Philippe Lopez, Cristian Lupu, Gabor Radnoti, David Richardson, Alberto Trevisan, Yann Tr\'emolet, Freja Vamborg, S\'ebastien Villaume, and Jean\textendash~No\"el Th\'epaut.
\newblock The era5 global reanalysis.
\newblock \emph{Quarterly Journal of the Royal Meteorological Society}, 146\penalty0 (730):\penalty0 1999--2049, 2020.
\newblock \doi{10.1002/qj.3803}.
\newblock URL \url{https://doi.org/10.1002/qj.3803}.

\bibitem[McNally et~al.(2024)McNally, Lessig, Lean, Boucher, Alexe, Pinnington, Chantry, Lang, Burrows, Chrust, Pinault, Villeneuve, Bormann, and Healy]{mcnally2024}
Anthony McNally, Christian Lessig, Peter Lean, Eulalie Boucher, Mihai Alexe, Ewan Pinnington, Matthew Chantry, Simon Lang, Chris Burrows, Marcin Chrust, Florian Pinault, Ethel Villeneuve, Niels Bormann, and Sean Healy.
\newblock Data driven weather forecasts trained and initialised directly from observations.
\newblock \emph{arXiv preprint arXiv:2407.15586}, 2024.
\newblock \doi{10.48550/arXiv.2407.15586}.
\newblock URL \url{https://arxiv.org/abs/2407.15586}.

\bibitem[Alexe et~al.(2024)Alexe, Boucher, Lean, Pinnington, Laloyaux, McNally, Lang, Chantry, Burrows, Chrust, Pinault, Villeneuve, Bormann, and Healy]{alexe2024}
Mihai Alexe, Eulalie Boucher, Peter Lean, Ewan Pinnington, Patrick Laloyaux, Anthony McNally, Simon Lang, Matthew Chantry, Chris Burrows, Marcin Chrust, Florian Pinault, Ethel Villeneuve, Niels Bormann, and Sean Healy.
\newblock {G}raph{DOP}: Towards skilful data-driven medium-range weather forecasts learnt and initialised directly from observations, 2024.
\newblock URL \url{https://arxiv.org/abs/2412.15687}.

\bibitem[Allen et~al.(2025)Allen, Markou, Tebbutt, Requeima, Bruinsma, Andersson, Herzog, Lane, Chantry, Hosking, and Turner]{Allen2025Aardvark}
Anna Allen, Stratis Markou, Will Tebbutt, James Requeima, Wessel~P. Bruinsma, Tom~R. Andersson, Michael Herzog, Nicholas~D. Lane, Matthew Chantry, J.~Scott Hosking, and Richard~E. Turner.
\newblock End-to-end data-driven weather prediction.
\newblock \emph{Nature}, 641\penalty0 (8065):\penalty0 1172--1179, 2025.
\newblock \doi{10.1038/s41586-025-08897-0}.
\newblock URL \url{https://doi.org/10.1038/s41586-025-08897-0}.

\bibitem[Vandal et~al.(2024)Vandal, Duffy, McDuff, Nachmany, and Hartshorn]{vandal2024}
Thomas~J. Vandal, Kate Duffy, Daniel McDuff, Yoni Nachmany, and Chris Hartshorn.
\newblock Global atmospheric data assimilation with multi-modal masked autoencoders, 2024.
\newblock URL \url{https://arxiv.org/abs/2407.11696}.

\bibitem[Keller and Potthast(2024)]{keller2024}
Jan~D. Keller and Roland Potthast.
\newblock Ai-based data assimilation: Learning the functional of analysis estimation, 2024.
\newblock URL \url{https://arxiv.org/abs/2406.00390}.

\bibitem[Sun et~al.(2024)Sun, Zhong, Xu, Huang, Li, Neelin, Chen, Feng, Han, Wu, and Qi]{sun2024}
Xiuyu Sun, Xiaohui Zhong, Xiaoze Xu, Yuanqing Huang, Hao Li, J.~David Neelin, Deliang Chen, Jie Feng, Wei Han, Libo Wu, and Yuan Qi.
\newblock Fuxi weather: A data-to-forecast machine learning system for global weather, 2024.
\newblock URL \url{https://arxiv.org/abs/2408.05472}.

\bibitem[Lean et~al.(2025)Lean, Alexe, Boucher, Pinnington, Lang, Laloyaux, Bormann, and McNally]{Lean25}
Peter Lean, Mihai Alexe, Eulalie Boucher, Ewan Pinnington, Simon Lang, Patrick Laloyaux, Niels Bormann, and Anthony McNally.
\newblock Learning from nature: insights into {G}raph{DOP}'s representations of the earth system, 2025.
\newblock URL \url{https://arxiv.org/abs/2508.18018}.

\bibitem[Boucher et~al.(2025)Boucher, Alexe, Lean, Pinnington, Lang, Laloyaux, Zampieri, de~Rosnay, Bormann, and McNally]{Boucher25}
Eulalie Boucher, Mihai Alexe, Peter Lean, Ewan Pinnington, Simon Lang, Patrick Laloyaux, Lorenzo Zampieri, Patricia de~Rosnay, Niels Bormann, and Anthony McNally.
\newblock Learning coupled earth system dynamics with graphdop, 2025.
\newblock URL \url{https://arxiv.org/abs/2510.20416}.

\bibitem[Malardel et~al.(2016)Malardel, Wedi, Deconinck, Diamantakis, Kuehnlein, Mozdzynski, Hamrud, and Smolarkiewicz]{malardel2016}
Sylvie Malardel, Nils Wedi, Willem Deconinck, Michail Diamantakis, Christian Kuehnlein, G.~Mozdzynski, M.~Hamrud, and P.~Smolarkiewicz.
\newblock A new grid for the {IFS}, 2016.
\newblock URL \url{https://www.ecmwf.int/node/17262}.

\bibitem[Kakkad et~al.(2023)Kakkad, Jannu, Sharma, Aggarwal, and Medya]{kakkad2023}
Jaykumar Kakkad, Jaspal Jannu, Kartik Sharma, Charu Aggarwal, and Sourav Medya.
\newblock A survey on explainability of graph neural networks, 2023.
\newblock URL \url{https://arxiv.org/abs/2306.01958}.

\bibitem[Chen et~al.(2024)Chen, Bian, Han, and Cheng]{chen2024}
Yongqiang Chen, Yatao Bian, Bo~Han, and James Cheng.
\newblock How interpretable are interpretable graph neural networks?, 2024.
\newblock URL \url{https://arxiv.org/abs/2406.07955}.

\bibitem[Ancell and Hakim(2007)]{Ancell07}
Brian Ancell and Gregory~J. Hakim.
\newblock Comparing adjoint- and ensemble-sensitivity analysis with applications to observation targeting.
\newblock \emph{Monthly Weather Review}, 135\penalty0 (12):\penalty0 4117 -- 4134, 2007.
\newblock \doi{10.1175/2007MWR1904.1}.

\bibitem[Zhu and Gelaro(2008)]{Zhu08}
Yanqiu Zhu and Ronald Gelaro.
\newblock Observation sensitivity calculations using the adjoint of the gridpoint statistical interpolation ({GSI}) analysis system.
\newblock \emph{Monthly Weather Review}, 136\penalty0 (1):\penalty0 335 -- 351, 2008.
\newblock \doi{10.1175/MWR3525.1}.

\bibitem[Lopez(2001)]{Lopez01}
Philippe Lopez.
\newblock The inclusion of {3D} prognostic cloud and precipitation variables in adjoint calculations.
\newblock \emph{Monthly Weather Review}, 131\penalty0 (9):\penalty0 1953--1974, 2001.

\bibitem[Errico et~al.(2003)Errico, Raeder, and Fillion]{Errico03}
Ronald~M. Errico, Kevin~D. Raeder, and Luc Fillion.
\newblock Examination of the sensitivity of forecast precipitation rates to possible perturbations of initial conditions.
\newblock \emph{Tellus A: Dynamic Meteorology and Oceanography}, 55\penalty0 (1):\penalty0 88--105, 2003.

\bibitem[Mahfouf and Bilodeau(2007)]{Mahfouf07}
Jean-François Mahfouf and Bernard Bilodeau.
\newblock Adjoint sensitivity of surface precipitation to initial conditions.
\newblock \emph{Monthly Weather Review}, 135\penalty0 (8):\penalty0 2879 -- 2896, 2007.
\newblock \doi{10.1175/MWR3439.1}.

\bibitem[Slivinski et~al.(2024)Slivinski, Whitaker, Frolov, Smith, and Agarwal]{slivinski2024}
Laura~C. Slivinski, Jeffrey~S. Whitaker, Sergey Frolov, Timothy~A. Smith, and Niraj Agarwal.
\newblock Assimilating observed surface pressure into ml weather prediction models, 2024.
\newblock URL \url{https://arxiv.org/abs/2412.18016}.

\bibitem[Tian et~al.(2024)Tian, Holdaway, and Kleist]{tian2024}
Xiaoxu Tian, Daniel Holdaway, and Daryl Kleist.
\newblock Exploring the use of machine learning weather models in data assimilation, 2024.
\newblock URL \url{https://arxiv.org/abs/2411.14677}.

\bibitem[Solvik et~al.(2025)Solvik, Penny, and Hoyer]{Solvik25}
Kylen Solvik, Stephen~G. Penny, and Stephan Hoyer.
\newblock {4D-Var} using hessian approximation and backpropagation applied to automatically differentiable numerical and machine learning models.
\newblock \emph{Journal of Advances in Modeling Earth Systems}, 17\penalty0 (4):\penalty0 e2024MS004608, 2025.
\newblock \doi{https://doi.org/10.1029/2024MS004608}.

\bibitem[Evensen(2003)]{Evensen03}
Geir Evensen.
\newblock The ensemble kalman filter: theoretical formulation and practical implementation.
\newblock \emph{Ocean Dynamics}, 53\penalty0 (4):\penalty0 343--367, 2003.

\bibitem[Isaksen et~al.(2010)Isaksen, Bonavita, Buizza, Fisher, Haseler, Leutbecher, and Raynaud]{isaksen2010eda}
L.~Isaksen, M.~Bonavita, R.~Buizza, M.~Fisher, J.~Haseler, M.~Leutbecher, and L.~Raynaud.
\newblock Ensemble of data assimilations at {ECMWF}.
\newblock {ECMWF} Technical Memorandum No. 636, 2010.

\bibitem[Hunt et~al.(2007)Hunt, Kostelich, and Szunyogh]{Hunt07}
Brian~R Hunt, Eric~J Kostelich, and Istvan Szunyogh.
\newblock Efficient data assimilation for spatiotemporal chaos: A local ensemble transform {K}alman filter.
\newblock \emph{Physica D: Nonlinear Phenomena}, 230\penalty0 (1-2):\penalty0 112--126, 2007.
\newblock \doi{10.1016/j.physd.2006.11.008}.

\bibitem[Chen and Oliver(2017)]{Yan17}
Yan Chen and Dean Oliver.
\newblock Localization and regularization for iterative ensemble smoothers.
\newblock \emph{Computational Geosciences}, 21\penalty0 (1):\penalty0 13--30, 2017.
\newblock \doi{10.1007/s10596-016-9599-7}.

\bibitem[Gaspari and Cohn(1999)]{Gaspari99}
Gregory Gaspari and Stephen~E. Cohn.
\newblock Construction of correlation functions in two and three dimensions.
\newblock \emph{Quarterly Journal of the Royal Meteorological Society}, 125\penalty0 (554):\penalty0 723--757, 1999.

\bibitem[Marcinkevičs and Vogt(2023)]{Marcin2023}
Ričards Marcinkevičs and Julia Vogt.
\newblock Interpretable and explainable machine learning: A methods-centric overview with concrete examples.
\newblock \emph{WIREs Data Mining and Knowledge Discovery}, 13\penalty0 (3):\penalty0 e1493, 2023.
\newblock \doi{https://doi.org/10.1002/widm.1493}.
\newblock URL \url{https://wires.onlinelibrary.wiley.com/doi/abs/10.1002/widm.1493}.

\bibitem[Yang et~al.(2024)Yang, Hu, Li, Mu, Yu, Xia, Li, Dasgupta, and Xiong]{yang2024}
Ruyi Yang, Jingyu Hu, Zihao Li, Jianli Mu, Tingzhao Yu, Jiangjiang Xia, Xuhong Li, Aritra Dasgupta, and Haoyi Xiong.
\newblock Interpretable machine learning for weather and climate prediction: A survey, 2024.
\newblock URL \url{https://arxiv.org/abs/2403.18864}.

\bibitem[Lundberg and Lee(2017)]{lundberg2017unified}
Scott Lundberg and Su-In Lee.
\newblock A unified approach to interpreting model predictions.
\newblock In \emph{Proceedings of Advances in Neural Information Processing Systems}, volume~30, 2017.

\bibitem[Ribeiro et~al.(2016)Ribeiro, Singh, and Guestrin]{ribeiro2016should}
Marco~Tulio Ribeiro, Sameer Singh, and Carlos Guestrin.
\newblock "{W}hy should {I} trust you?": Explaining the predictions of any classifier, 2016.
\newblock URL \url{https://arxiv.org/abs/1602.04938}.

\bibitem[Lorenc and Marriott(2014)]{Lorenc14}
Andrew Lorenc and Richard Marriott.
\newblock Forecast sensitivity to observations in the met office global numerical weather prediction system.
\newblock \emph{Quarterly Journal of the Royal Meteorological Society}, 140\penalty0 (678):\penalty0 209--224, 2014.
\newblock \doi{10.1002/qj.2122}.

\bibitem[Errico(2007)]{Errico07}
Ronald Errico.
\newblock Interpretations of an adjoint-derived observational impact measure.
\newblock \emph{Tellus A: Dynamic Meteorology and Oceanography}, 59\penalty0 (2):\penalty0 273--276, 2007.
\newblock \doi{10.1111/j.1600-0870.2006.00217.x}.

\bibitem[McNally(2019)]{Mcnally19}
Anthony McNally.
\newblock On the sensitivity of a {4D-Var} analysis system to satellite observations located at different times within the assimilation window.
\newblock \emph{Quarterly Journal of the Royal Meteorological Society}, 145\penalty0 (723):\penalty0 2806--2816, 2019.
\newblock \doi{10.1002/qj.3596}.

\bibitem[Liu and Kalnay(2008)]{Liu08}
Junjie Liu and Eugenia Kalnay.
\newblock Estimating observation impact without adjoint model in an ensemble {K}alman filter.
\newblock \emph{Quarterly Journal of the Royal Meteorological Society}, 134\penalty0 (634):\penalty0 1327--1335, 2008.
\newblock \doi{https://doi.org/10.1002/qj.280}.

\bibitem[Necker et~al.(2018)Necker, Weissmann, and Sommer]{Necker18}
Tobias Necker, Martin Weissmann, and Matthias Sommer.
\newblock The importance of appropriate verification metrics for the assessment of observation impact in a convection-permitting modelling system.
\newblock \emph{Quarterly Journal of the Royal Meteorological Society}, 144\penalty0 (714):\penalty0 1667--1680, 2018.
\newblock \doi{https://doi.org/10.1002/qj.3390}.

\bibitem[Prive et~al.(2021)Prive, Errico, Todling, and El~Akkraoui]{Prive21}
N.C. Prive, Ronald~M. Errico, Ricardo Todling, and Amal El~Akkraoui.
\newblock Evaluation of adjoint-based observation impacts as a function of forecast length using an observing system simulation experiment.
\newblock \emph{Quarterly Journal of the Royal Meteorological Society}, 147\penalty0 (734):\penalty0 121--138, 2021.
\newblock \doi{https://doi.org/10.1002/qj.3909}.

\bibitem[Hotta et~al.(2017)Hotta, Chen, Kalnay, Ota, and Miyoshi]{Daisuke17}
Daisuke Hotta, Tse-Chun Chen, Eugenia Kalnay, Yoichiro Ota, and Takemasa Miyoshi.
\newblock Proactive qc: A fully flow-dependent quality control scheme based on {EFSO}.
\newblock \emph{Monthly Weather Review}, 145\penalty0 (8):\penalty0 3331 -- 3354, 2017.
\newblock \doi{10.1175/MWR-D-16-0290.1}.

\bibitem[Chen and Kalnay(2020)]{Tse20}
Tse-Chun Chen and Eugenia Kalnay.
\newblock Proactive quality control: Observing system experiments using the {NCEP} global forecast system.
\newblock \emph{Monthly Weather Review}, 148\penalty0 (9):\penalty0 3911 -- 3931, 2020.
\newblock \doi{10.1175/MWR-D-20-0001.1}.

\end{thebibliography}

\newpage
\appendix

\section*{Appendix}

\section{Specification of training datasets}

\begin{table}[htbp]
\centering
\begin{tabular}{|l|l|c|l|}
\hline
\textbf{Category} & \textbf{Instrument} & \textbf{Period} & \textbf{Parameters} \\
\hline
\multirow{6}{*}{Microwave Sounders} 
& NPP ATMS & 2013-2022 & surface channels 1-4, 16, 17\\ 
& & & temperature sounding channels 5-15\\ 
& & & water vapour channels 18-22\\
& NOAA 20 ATMS & 2018-2022 & surface channels 1-4, 16, 17\\ 
& & & temperature sounding channels 5-15\\ 
& & & water vapour channels 18-22\\ 
\hline
\multirow{3}{*}{Infrared Sounders}
& METOP-B IASI & 2013-2022 & temperature sounding channels 1-10  \\
& & & surface channels 11-13  \\
& & & water vapour channels 14-18  \\
\hline
\multirow{10}{*}{Conventional - surface}
& Automatic Land SYNOP & 2013-2023 & ps, t2m, rh2m, u10, v10 \\
& Manual Land SYNOP & 2013-2023 & ps, t2m, rh2m, u10, v10 \\
& BUFR Land SYNOP & 2014-2023 & ps, t2m, rh2m, u10, v10 \\
& SHIP & 2013-2023 & ps, t2m, rh2m, u10, v10 \\
& BUFR SHIP SYNOP & 2014-2023 & ps, t2m, rh2m, u10, v10 \\
& Abbreviated SHIP & 2013-2023 & ps, t2m, rh2m, u10, v10 \\
& METAR & 2013-2023 & ps, t2m, rh2m, u10, v10 \\
& Automatic METAR & 2013-2023 & ps, t2m, rh2m, u10, v10 \\
& DRIBU & 2013-2023 & ps, sst \\
& BUFR Drifting Buoys & 2016-2023 & ps, sst \\
\hline
\multirow{5}{*}{Conventional - sonde}
& TEMP SHIP & 2013-2022 & z, t, u, v on standard pressure levels \\
& BUFR SHIP TEMP & 2014-2022 & z, t, u, v on standard pressure levels \\
& Land TEMP & 2013-2022 & z, t, u, v on standard pressure levels \\
& BUFR Land TEMP & 2014-2022 & z, t, u, v on standard pressure levels \\ 
& Dropsondes & 2013-2022 & z, t, u, v on standard pressure levels \\
\hline
\end{tabular}
\caption{Input and output observations used in this study to train GraphDOP.}
\label{table:dop-observations}
\end{table}

\end{document}